\documentclass[12pt]{article}

\title{\begin{flushright}
{\normalsize McGill/96-37\\
NUC-MINN-96/16-T\\}
\end{flushright}
\vspace*{0.3in}
{\bf $\omega$ - $\phi$ mixing at finite temperature} }
\author{{Charles Gale}$^{\dag}$ and {David Seibert}$^{\ddag}$
 \vspace*{0.1in}\\
 {\it Physics Department, McGill University}\\ \vspace*{0.2in}
 {\it Montr\'eal, Qu\'ebec H3A 2T8, Canada}\\ \vspace*{0.1in}
{Joseph Kapusta}$^{\S}$ \\
  {\it School of Physics and Astronomy, University of Minnesota}\\
  {\it Minneapolis, MN 55455, USA}}

\date{}

\usepackage[dvips]{graphicx}
\begin{document}

\maketitle

\begin{center}
Abstract
\end{center}

\noindent
We compute the mass shifts and mixing of the $\omega$ and $\phi$
mesons at finite temperature due to scattering from thermal pions.
The $\rho$ and $b_1$ mesons are important intermediate states.
Up to a temperature of 140 MeV the $\omega$ mass increases by
12 MeV and the $\phi$ mass decreases by 0.6 MeV.
The change in mixing angles is negligible.

\vspace{1.in}

\noindent
$^{\dag}$ Electronic address: gale@hep.physics.mcgill.ca \\
\noindent
$^{\ddag}$ Permanent address: SoftQuad Inc., 108-10070 King George
Highway,\\ \indent Surrey, B.C. V3T 2W4, Canada.  Electronic address: 
seibert@direct.ca\\
\noindent
$^{\S}$ Electronic address: kapusta@physics.spa.umn.edu

\newpage

The emission of lepton pairs from high temperature matter formed
in high energy nucleus$-$nucleus collisions is of great theoretical
and experimental interest.  It can signal a change in the properties of
hadrons (more precisely, correlation functions) in hot hadronic matter
as it approaches a chiral symmetry or a quark deconfinement phase
transition or rapid crossover.  Lepton pair emission has been measured
by the DLS collaboration \cite{dls} at the (disassembled) Bevalac at LBNL,
by the CERES \cite{ceres} and HELIOS \cite{helios} collaborations
at the SPS at CERN, and will be measured in the PHENIX experiment \cite{cdr}
at RHIC at BNL.

Vector mesons are prominant in these studies because of their
coupling to the electromagnetic current.  Most studies have focussed
on the lighter ones, $\rho$, $\omega$, and $\phi$, and on the heavier
J/$\psi$.  The best signals are those that have a characteristic shape
or structure.  The $\rho$ meson is very broad in vacuum and undoubtedly
gets even broader at finite temperature.  The J/$\psi$ is narrow; it
has a whole literature of its own within the field.  The $\omega$ and
$\phi$ mesons are rather narrow, but not so narrow that they will
all decay after the hot matter has blown apart in a high energy collision.
This makes them good candidates to study.

In vacuum the $\phi$ meson is almost entirely $\bar{s}s$ in its valence
quark content while the $\omega$ meson is almost entirely nonstrange.
There is a small mixing as evidenced by the observed decay mode
$\phi \rightarrow \pi \rho$ and by studies using effective hadronic
Lagrangians.  In this paper we will study the change in masses and
mixing angles of the $\omega$ and $\phi$ at moderate
temperatures using only conventional ideas.  We consider our study
a simple extrapolation of known physics which may help in deciphering
future experiments.  Specifically, we study the scattering of these
mesons from thermal pions which are the most abundant mesons at
temperatures below 100 MeV or so.  We use effective Lagrangian
techniques to model the relatively soft interactions coupling
pions, $\omega$, and $\phi$ mesons, fitting the parameters to known
physical quantities.  The resulting scattering amplitudes are
used in a virial expansion to compute the vector meson properties
at finite temperature.

A survey of the literature on effective hadronic Lagrangians and
the Review of Particle Physics \cite{revpp} suggests that the
interactions of relevance
involve the 3--point vertices $\phi \rho \pi$, $\omega \rho \pi$,
$\phi b_1 \pi$, and $\omega b_1 \pi$.  The vector self$-$energies
are obtained by computing one loop diagrams at finite temperature.
We only keep the contribution from thermal pions in the loop.
This corresponds to a virial expansion where the self$-$energy is
obtained from the forward scattering amplitude of the vector meson
from a pion with a Bose$-$Einstein momentum distribution.  We do
not consider thermal scattering from kaons or $\eta$ mesons \cite{song}.
Those contributions would be important at higher temperatures and are
the same order in the Boltzmann factor as the scattering from
two thermal pions consecutively.  The latter would require the
computation of two loop self$-$energy diagrams at finite temperature
which is notoriously difficult.

The interaction involving the $\rho$ meson is described by the
Wess--Zumino term \cite{WZ}.
\begin{equation}
{\cal L}_{(\omega,\phi) \rho \pi} = g \, \epsilon^{\alpha \beta
\mu \nu} \partial_{\alpha} \mbox{\boldmath $\rho$}_{\beta} \cdot
\mbox{\boldmath $\pi$} \left( \frac{\partial_{\mu} \omega_{\nu}^8
+ \sqrt{2} \, \partial_{\mu} \omega_{\nu}^s}{\sqrt{3}} \right)
\end{equation}
The octet and singlet fields, $\omega_8$ and $\omega_s$,
are expressed in terms of the (vacuum)
physical fields with a mixing angle $\theta_V$.
\begin{eqnarray}
\omega_8 &=& \phi \cos \theta_V + \omega \sin \theta_V \nonumber \\
\omega_s &=& \omega \cos \theta_V - \phi \sin \theta_V
\end{eqnarray}
Ideal mixing is defined such that the physical $\phi$ meson would not
couple to nonstrange hadrons.  This corresponds to $\theta_V =
\theta_{\rm ideal} = \tan^{-1}(1/\sqrt{2}) \approx 35.3^\circ$.  The real
world is not far from that.  Durso \cite{durso}, for example,
fits 39.2$^\circ$, while the Review of Particle Physics quotes 39$^\circ$
based on the Gell--Mann Okubo mass formula.
The coupling constant $g$ is related to the coupling
$g_{VVP}$ used by Gomm, Kaymakcalan, and Schechter \cite{gks} by
$g = - \sqrt{2} \, g_{VVP}$.
Gauging the Wess--Zumino term yields the prediction \cite{gauged}
\begin{equation}
g = \frac{3 g_{\rho\pi\pi}^2}{8 \pi^2 f_{\pi}^2} \, .
\end{equation}
This is consistent with phenomenological studies.  For example, Durso
fits $g^2 = 1.62 \pm 0.19 \times 10^{-4}$ MeV$^{-2}$ with a
pseudoscalar mixing angle $\theta_P$ = $- 9.7^\circ$.
We will insure that the parameters chosen reproduce the
measured decay rate \cite{revpp} using
\begin{equation}
\Gamma_{\phi \rightarrow \rho\pi} = \frac{(\cos \theta_V - \sqrt{2}\,
\sin \theta_V)^2}{288 \pi} \frac{g^2}{m_{\phi}^3}
\left[ (m_{\phi}^2 + m_{\pi}^2 - m_{\rho}^2)^2 -4 m_{\pi}^2
m_{\phi}^2 \right]^{3/2} \, .
\end{equation}
Accepting Durso's value of $g$ we fit $\theta_V = 40.1^\circ$.

The $b_1(1235)$ has a branching ratio of more than 50\% into
$\omega \pi$ and less than 1.5\% into $\phi \pi$.  Thus, the $b_1$
meson is important for the mass shift of the $\omega$ meson at
finite temperature, as noticed by Shuryak \cite{Ed}.  The
interactions of this resonant meson is not much studied, hence we are
obliged to do so here.  The interactions are assumed to be SU(3) symmetric
with SU(3) broken only by mass terms.  Suppressing Lorentz indices
the interaction is of the form ${\bf b} \cdot \mbox{\boldmath $\pi$}
(\omega_8 + \sqrt{2} \omega_s)/\sqrt{3}$.  Suppressing isospin
indices there are five possibilities: $b_{\mu} \omega^{\mu} \pi$,
$b_{\mu \nu} \omega^{\mu \nu} \pi$, $b_{\mu} \omega^{\mu \nu}
\partial_{\nu} \pi$, $b_{\mu \nu} \omega^{\mu} \partial^{\nu} \pi$,
$b_{\mu} \omega_{\nu} \partial^{\mu} \partial^{\nu} \pi$.
The last of these forms is not independent of the second
one if it is integrated by parts and the field conditions
$\partial \cdot b = \partial \cdot \omega = 0$ are used.
The third and fourth forms are not independent either in the
weak field limit.  For example, the form $b_{\mu} \omega^{\mu \nu}
\partial_{\nu} \pi$ becomes $-\partial_{\nu} b_{\mu} \omega^{\mu\nu}
\pi - b_{\mu}\partial_{\nu} \omega^{\mu\nu} \pi$ upon integration
by parts.  The equation of motion is $\partial_{\mu}\omega^{\mu\nu}
= m_{\omega}^2 \omega^{\nu}$ + nonlinear field terms.  The term linear
in the $\omega$ field does not give rise to a new interaction term,
and the term which is nonlinear in the fields is not of relevance here.
Therefore the interaction Lagrangian is
\begin{equation}
{\cal L}_{(\omega,\phi)b_1 \pi} = g_{b_1} \, \mbox{\boldmath $\pi$}
\cdot {\bf b}^{\mu} \left( \frac{ \omega_{\mu}^8
+ \sqrt{2} \, \omega_{\mu}^s}{\sqrt{3}} \right) +
h_{b_1} \, \mbox{\boldmath $\pi$}
\cdot {\bf b}^{\mu\nu} \left( \frac{ \omega_{\mu\nu}^8
+ \sqrt{2} \, \omega_{\mu\nu}^s}{\sqrt{3}} \right) \, .
\end{equation}

The two coupling constants can be inferred from the decay rate
$b_1 \rightarrow \omega \pi$ and from the ratio of the D wave content of
the decay amplitude to its S wave content.
The spin-averaged squared matrix element for the decay rate is
\begin{eqnarray}
| \overline{{\cal M}}_{b_1 \rightarrow \omega \pi}|^2 &=&
( \sin\theta_V + \sqrt{2} \cos\theta_V)^2  \left\{
g_{b_1}^2 \left[ 2 + {{( k \cdot q )^2}\over {m_{b_1}^2 m_{\omega}^2}}
\right]\;+ \;12 g_{b_1} h_{b_1} ( k \cdot q )  \right. \nonumber \\
&+& \left. \phantom{\frac{1}{1}}
4 h_{b_1}^2 \left[
m_{b_1}^2 m_{\omega}^2 + 2\; ( k \cdot q )^2 \right] \right\} \, ,
\label{eqm2}
\end{eqnarray}
where $k \cdot q$ is the scalar product of the $b_1$ and $\omega$
four-momenta.  By expanding the decay amplitude in spherical harmonics
and comparing this expansion with an expression of the total decay amplitude
containing the actual helicities, we relate the individual $S$ and $D$
amplitudes to the coupling constants as
\begin{eqnarray}
f^S &=& {\sqrt{4 \pi}\over 3}  \left\{  \left[ g_{b_1} + 2 h_{b_1}
(k\cdot q ) \right]
 {2 m_\omega + E_\omega \over m_\omega} - 2 h_{b_1} |\, {\bf q}\, |^2
{m_{b_1}\over m_\omega} \right\} \, , \nonumber \\
f^D &=& {\sqrt{8 \pi}\over 3}  \left\{ \left[ g_{b_1} + 2 h_{b_1}
(k\cdot q ) \right]
 {m_\omega - E_\omega \over m_\omega} + 2 h_{b_1} |\, {\bf q}\, |^2
{m_{b_1}\over m_\omega} \right\} \, .
\end{eqnarray}
Here $E_\omega$ and $\bf{q}$ are the energy and momentum of
the $\omega$ in the $b_1$ rest frame.
The Review of Particle Physics gives the width 142$\pm$8 MeV and the
measured D/S ratio 0.26$\pm$0.04.  Fitting to central values determines
the most likely values of the coupling constants to be
$g_{b_1} = -9.471$ GeV and $h_{b_1} = 6.642$ GeV$^{-1}$.

The location of the poles and the mixing angle at finite temperature
are obtained by finding the zeros of the inverse propagator in the
(vacuum) physical $\omega$--$\phi$ basis.
\begin{equation}
{\cal D}^{-1}(k_0,{\bf k}) = k_0^2 - {\bf k}^2 - M^2 - \Sigma(k_0,{\bf k})
\end{equation}
$M^2$ is the 2$\times$2 mass matrix at zero temperature.
It is diagonal with components $M^2_{11} = m_{\omega}^2$ and
$M^2_{22} = m_{\phi}^2$.  The self$-$energy has contributions from
the $\rho$ meson and from the $b_1$ meson.  The first was calculated
by Haglin and Gale \cite{HG}.  The second is readily calculated by the usual
finite temperature rules \cite{Kap}.  We restrict our attention to the finite
temperature contribution.  We consider only $\omega$ and $\phi$ mesons
at rest in the many--body system (${\bf k} = 0$) since that is
where temperature will have its maximum impact.  As $|{\bf k}|$
increases, many$-$body effects will decrease, and in the limit
$|{\bf k}| \gg T$ they will disappear altogether.
Finally, the imaginary part of the self$-$energy is small compared to
the real part (inclusive of $M^2$) and so we do not include it in
our present calculation.  We find
\begin{eqnarray}
\Sigma(k_0,|{\bf k}|=0) = \left( \begin{array}{cc}
\cos^2\delta_V & - \left(\cos\delta_V \sin\delta_V \right)/2 \\
- \left(\cos\delta_V \sin\delta_V \right)/2 & \sin^2\delta_V
\end{array} \right)
\left( \Sigma_{\rho} + \Sigma_{b_1} \right) \, ,
\end{eqnarray}
where $\delta_V = \theta_V - \theta_{\rm ideal}$ measures the
deviation from ideal mixing. The scalar functions are
\begin{equation}
\Sigma_{\rho} = - \frac{g^2 k_0^2 (k_0^2 +m_{\pi}^2 - m_{\rho}^2)}
{\pi^2} \int_0^{\infty} \frac{dp \, p^4}{\omega_{\pi}}
\frac{1}{\exp(\omega_{\pi}/T) - 1} \cdot
\frac{1}{(k_0^2 +m_{\pi}^2 - m_{\rho}^2)^2 - 4 k_0^2 \omega_{\pi}^2}
\end{equation}
and
\begin{figure}
\begin{center}
\includegraphics[angle=90, width=8cm]{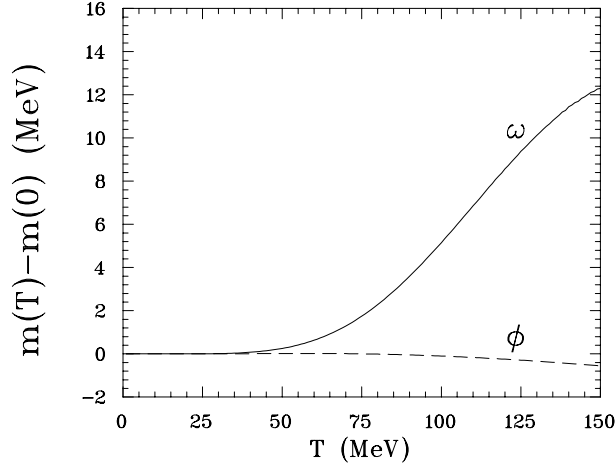}
\end{center}
\caption{\small  Mass shifts of the $\omega$ and $\phi$ 
mesons as functions
of temperature due to scattering from thermal pions.  Multiple pion
scattering and scattering from other thermal mesons should become
important above 100 MeV temperature.} 
\end{figure}
\begin{eqnarray}
\Sigma_{b_1} &=& - \frac{1} {2\pi^2 m_{b_1}^2}
\int_0^{\infty} \frac{dp \, p^2}{\omega_{\pi}}
\frac{1}{\exp(\omega_{\pi}/T) - 1} \cdot
\frac{1}{(k_0^2 +m_{\pi}^2 - m_{b_1}^2)^2 - 4 k_0^2 \omega_{\pi}^2}\
 \nonumber \\ &\times&
\left\{\left( k_0^2 + m_\pi^2 - m_{b_1}^2 \right) \left( 3 m_{b_1}^2 \left[
g_{b_1}^2 + 4 h_{b_1}^2 k_0^2 \left( k_0^2 - m_\pi^2 \right) + 4 g_{b_1}
h_{b_1} k_0^2 \right] + g_{b_1}^2 p^2 \right) \right. \nonumber \\
&+&  8 h_{b_1}^2 m_{b_1}^2 k_0^2 \left[ 3 m_\pi^2 \left( m_\pi^2 -
k_0^2 - m_{b_1}^2 \right) + p^2 \left( m_\pi^2 - m_{b_1}^2 - 5 k_0^2 \right)
\right] \nonumber \\
&-& \left. 24 g_{b_1} h_{b_1} k_0^2 m_{b_1}^2 \omega_\pi^2 \right\}\ ,
\end{eqnarray}
where $\omega_{\pi} = \sqrt{p^2 + m_{\pi}^2}$.
Even in the limit of zero pion mass these integrals cannot be
evaluated exactly in terms of elementary functions and so
we resort to numerical integration.  It should be remarked that
both integrals can have a pole within the limits of integration
depending on the value of $k_0$.  This is handled by the principal
value prescription; the imaginary parts are not displayed separately.
Inclusion of a finite width for both the intermediate $\rho$ and
$b_1$ mesons or, better yet, use of dressed vacuum propagators,
would lead to finite integrals.  Only if we find a signficant
many$-$body effect should it be necessary to include this next level
of sophistication.

After diagonalization one finds the mass shifts
as functions of the temperature.  They are displayed in
Fig. 1.  Although the results are plotted
up to a temperature of 150 MeV to see the effect, other scattering
processes will come into play above 100 MeV, as mentioned earlier.
The $\omega$ mass goes up and a shift of roughly 12 MeV is reached at a
temperature of 140 MeV. The $\phi$ mass decreases monotonically
to about 0.6 MeV below its vacuum value at a temperature of 140 MeV.

Although a
rigorous exploration of the available parameter space is not the point
of this work, one can test the effect of simple variations. We
have thus chosen a combination that enhances the mass shift of the
$\omega$.  The values $g_{b_1} = -11.401$ GeV and
$h_{b_1} = 7.647$ GeV$^{-1}$
correspond to $D/S$ = 0.3 and $\Gamma_{b_1 \rightarrow \omega \pi}$ =
150 MeV. We are therefore still inside the measured error bars of
the $b_1$'s hadronic properties. This choice of couplings represents our
``maximum effect'' set. The result of this exercise is to increase
the mass shift of the $\omega$ and the $\phi$ by about 30\% and 25\%,
respectively. This illustrates the sensitivity of our effective
Lagrangian approach to empirical hadronic properties.
\begin{figure}
\begin{center}
\includegraphics[angle=90, width=8cm]{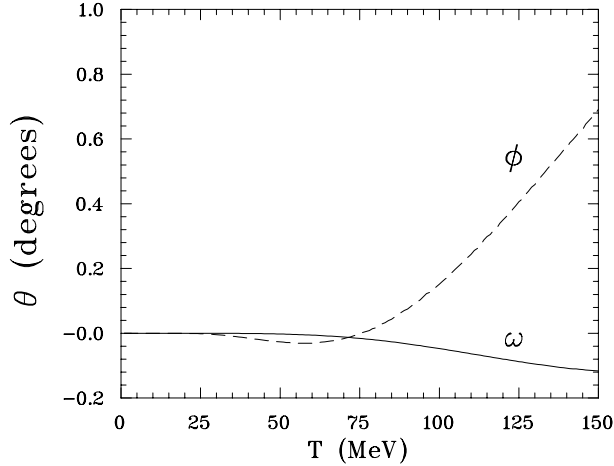}
\end{center}
\caption{\small  Mixing angles of the $\omega$ and $\phi$ mesons as
functions of temperature.  Same remarks as Fig. 1.}   
\end{figure}

The mixing angles at finite temperature are shown in Fig. 2.
The mixing angles remain uninterestingly small at all temperatures.

In conclusion, we have computed the change in the masses of the
$\omega$ and $\phi$ mesons and their mixing angle up to temperatures
of 100 MeV (pushing 150 MeV with the above caveat).  The calculation
is based on conventional physical input.  The change
in the $\phi$ mass and mixing angles are totally negligible.
The change in the $\omega$ mass is large enough to be potentially
detectable in future heavy ion experiments.

\section*{Acknowledgements}

One of us (C. G.) is happy to thank S. Gao for useful contributions. 
This work was supported in part by by the Natural Sciences and
Engineering Research Council of Canada, in part by the FCAR fund of the
Qu\'ebec Government, and in part by the US Department of Energy
under grant DE-FG02-87ER40328.

\end{document}